\documentclass[aps,prc,twocolumn,showpacs]{revtex4}
\usepackage{graphicx} 
\usepackage{dcolumn} 
\usepackage{bm}         
\begin{document}

\title{Longitudinal Fluctuations in Partonic and Hadronic Initial State}

\author{
Yun Cheng$^1$, Yu-Liang Yan$^2$, Dai-Mei Zhou$^1$, Xu Cai$^1$, Ben-Hao Sa$^2$,
and   Laszlo P. Csernai$^{3,4}$}

\affiliation{
$^1$
Institute of Particle Physics,
Huazhong Normal University,
430079 Wuhan, China\\
$^2$
China Institute of Atomic Energy,
P. O. Box 275 (18), 102413 Beijing, China\\
$^3$
Department of Physics and Technology,
University of Bergen, 5007, Bergen, Norway\\
$^4$
MTA-KFKI Research Inst for Particle and Nuclear Physics,
1525 Budapest, Pf. 49, Hungary
}
\date{\today}

\begin{abstract}
Collective flow in collisions between Lead nuclei at LHC are influenced by
random initial state fluctuations, especially for odd harmonics. Here we extend
fluctuation studies to longitudinal fluctuations, which may have significant effect
on the rapidity distribution of odd harmonics. Furthermore center of mass
rapidity fluctuations are measurable, but not yet analysed. Here in the PACIAE
parton and hadron molecular dynamics model we make an analysis of initial state
fluctuations. As previous analyses discussed mainly the effects of fluctuations
on eccentricity and the elliptic flow we pay particular attention to the
fluctuations of the Center of Mass rapidity of the system, which is conservatively estimated in our model as $\Delta y_{CM} = 0.1$,
by neglecting all pre-equilibrium emission effects that are increasing
the $y_{CM}$ fluctuations.
\end{abstract}


\maketitle

\section{ Introduction }

Global collective observables are becoming the most essential observables
in ultra-relativistic heavy ion reactions%
\cite{ALICE-Flow1}.
When we want to extract
precise knowledge from experiments, both on the Equation of State
(EoS) and the transport properties of matter \cite{Son,CKM}
we have to invoke
a most realistic description with fully 3+1 dimensional dynamical
evolution at all stages of the reaction, including the initial
state. This most adequate description of all stages can only be
achieved by the multi-module, hybrid models.
(See e.g. refs. \cite{Hybrid}.)

The initial state, where we have very little direct experimental information,
is of paramount importance in the theoretical description.
This leads to a wide variety of initial state models, which behave
differently. Theoretical models and experimental results indicate that
the initial state fluctuations are essential in understanding the
data, although in the global continuum  (fluid dynamical or field theoretical)
models these fluctuation effects may inherently not be present and
even may not survive to the hadronic final state.
Nevertheless, we need to analyze the behavior of these initial state
models from the point of view of fluctuations.
(See e.g. refs. \cite{Fluc}.)

However, one has to take into account that the Center of Mass (CM)
rapidity is not exactly the
same for all events because of random fluctuations in the
initial state caused by the difference of participant nucleon numbers
from projectile and target. This leads to considerable fluctuations at
large impact parameters, where the flow asymmetry is the
strongest, but the number of participant nucleons is the smallest.

Just as all initial
state fluctuations, we have two sources of CM-rapidity fluctuations:
 First, the number of nucleons are randomly located in the configuration
space and due to their fluctuating location, the number of participants
from the target and projectile nucleus must not be the same
event-by-event, even in the symmetric, A+A, collisions.
 Second, those nucleons, which are in the geometrical
participant zone, may actually not collide with any single nucleon
from the opposite nucleus, consequently these will not become participants.
Some recent results on the subject concerning the $v_2$ and $v_3$
fluctuations are discussed in refs. \cite{Qin10,Pet11,Grassi}.

Up to now less attention is paid to the fluctuations in the
beam direction. The expected momentum and/or rapidity fluctuations
in this direction may be bigger due to the large beam momentum
in recent experiments.
In case of CM-rapidity fluctuations there is an additional problem:
It is not obvious how tightly bound system is the initial state.
The number of participant nucleons may not come from the projectile and
the target nuclei equally, there can be
one or a few more nucleons from one side.
The momentum carried by the extra nucleons, may be shared (i) by
all participants equally in a tightly bound system (a single large
confined QGP bag, may be considered as such a system) or (ii) by a loosely
connected cloud of nucleons (where the extra nucleons have little
direct effect on the participant matter). In the later case, although
the total momentum is conserved, the internal energy of the participant
matter is increased considerably by the energy of the extra nucleons but
the momentum of the participant matter is not correlated with the momenta of
the extra nucleons. So, the collective rapidity change is much less.

It is important to mention that the phase transitions and the consequent
fluctuations both in and out of QGP may enhance the collective
behavior of the system \cite{CK92}. However, it is rather difficult to
estimate the consequences of such transitions and fluctuations to the
CM-rapidity fluctuations. From the point of view of initial state
fluctuations we have to arrive at system, which is close to local
equilibrium, thus, at high energies the transition to QGP has
to happen earlier than the formation of the initial state.
Thus, it is important to study the CM-rapidity fluctuation as an
observable on its own to learn about energy deposition, and also due
to its strong effect on flow observables. (See e.g. ref. \cite{CMSS}).

In this work, after some simple considerations, we present an analysis
of these fluctuations in the PACIAE model, where the major sources of
fluctuations are taken into account.
\vskip -1mm

\section{Analytical estimates for the CM-rapidity fluctuations}

As mentioned above, the initial state fluctuation is stemming from the
participant nucleon number ($N\!a + N\!b=N_{part}$) fluctuation.
Here $N\!a$ and $N\!b$ are the numbers of participant nucleons
from the projectile and target nuclei, respectively.
The participant matter forms then the initial state
system. In the following examples we present three
situations where different fraction of the beam energy is contributing
to the total transverse mass of the locally equilibrated participant
matter.

Let us first estimate the effect of fluctuations of the participant
matter for a impact parameter of
 $b=0.7b_{max}$ collision in
 Pb+Pb reactions at the
 LHC energy of $1.38 + 1.38$ A$\cdot$TeV,
for a tightly bound and unexcited system.
We assume that one extra nucleon from the projectile
nucleus will be absorbed into the participant
matter, which otherwise would contain
$N_{part}=N\!a{+}N\!b=32.5{+}32.5 = 65$
nucleons. Then this extra projectile nucleon,
$\delta N \equiv N\!a {-} N\!b=1$, carries
$m_t * sinh(y_0)$ momentum,
where $y_0 = 8$ is the beam rapidity at the above LHC energy and
$m_t = m_N$ is the transverse mass of a nucleon in the beam.
If this extra
momentum is absorbed in the participant matter, then according to the
momentum conservations:
\begin{eqnarray}
P_z = M^{CM}_t \sinh (\bigtriangleup y_{CM}) &=& \delta N \ m_t \sinh (y_0) \,
\label{Pzc}\\
  E = M^{CM}_t \cosh (\bigtriangleup y_{CM}) &=& N_{part}
\  m_t \cosh (y_0) \,
\label{Ec}
\end{eqnarray}
this extra nucleon will lead to a change
of the CM-rapidity, $\bigtriangleup y_{CM}$ (which is zero if the participant
nucleons are coming in equal numbers from the projectile and target). In the
above equations the $M^{CM}_t$ is transverse mass of the participant matter.

In the initial state model based on expanding flux tubes or streaks
\cite{M2001-2} used in fluid dynamical
calculations
\cite{CMSS,hydro1},
the initial state system is tightly bound and stopped within each
``streak". Thus, this model is applicable streak by streak and
its momentum change is more pronounced for the peripheral streaks
where the asymmetry
between the projectile and target involvements is the biggest.
In this initial state model
the transverse mass,
$M^{CM}_t$ is more than what would arise from the nucleon masses,
$ N_{part} m_N$, due to the field strength in the string. So
$ M^{CM}_t = N_{part}(m_t + L \sigma$),
where $L$ is the length of the streak and $\sigma$ is the effective
string tension. If the participant matter is
weakly excited, $M^{CM}_t \approx N_{part} (m_t + 1 {\rm GeV}) $.
The resulting shift of CM-rapidity can be derived from Eq. (\ref{Pzc}):
$$
\bigtriangleup y_{CM} \approx
{\rm arsinh} \left[ \frac{\delta N \ m_t}{N_{part} (m_t + 1{\rm GeV })}
\sinh (y_0)  \right] = 3.1 \, .
$$
Thus, CM-rapidity fluctuations may be quite substantial. In this case a large
fraction of beam energy should be carried away through other channels,
like pre-equilibrium emission.

For the initial state in hadronic transport models the momentum of extra
nucleons are hardly influencing the momenta of the other participant
nucleons. The extra nucleons are not stopped in this picture, the
transverse mass ($M^{CM}_t$) in the above expression includes large
pre-thermal momenta, but $M^{CM}_t$ can still be proportional to
$m_t * sinh(y_0)$. In such a model
the CM-rapidity fluctuation will be significantly smaller. For example, in
the above $b=0.7b_{max}$ Pb+Pb reaction at (1.38+1.38) A$\cdot$TeV if we
assume $65+\delta N$, (where $\delta N=1$) participant nucleons and
full equilibration, so that 2/3$^{rd}$ of the beam kinetic energy is
converted into the transverse mass of the participant matter, and
$M^{CM}_t$ can be approximated as
$M^{CM}_t=N_{part}(m_t+\epsilon_0 * 2/3)$ where $\epsilon_0 = 1.38$TeV
per nucleon in the Lab/CM frame. Then
the CM-rapidity fluctuation can be approximated as
\begin{equation}
\bigtriangleup y_{CM} \approx
{\rm arsinh} \left[ \frac{\delta N \ m_t}{N_{part} (m_t {+} 2\epsilon_0 /3)}
\sinh (y_0)  \right] = 0.025 \, .
\end{equation}
Although here we discuss the hadronic initial state in a
hadronic transport model, it is suitable for the partonic initial state
in hadron and parton transport models also.

The other limiting case is when all reaction energy is absorbed in the
participant matter. Then both Eqs. (\ref{Pzc},\ref{Ec}) are satisfied, and
for the same example of Pb+Pb collision as above the resulting CM-rapidity is
\begin{equation}
\bigtriangleup y_{CM} =
{\rm artanh} \left[ \frac{\delta N}{N_{part}} \tanh (y_0) \right]=0.015 \ .
\label{emom1}
\end{equation}

The above considerations show that the question of initial state
fluctuations is a rather complex and model dependent question.
After all, the collectivity or looseness of the initial state must be
estimated experimentally. The CM-rapidity fluctuations may provide a
very good tool to this research.

\section{Longitudinal Fluctuations in Partonic Initial State in PACIAE model}

We discussed above the hadronic initial state, now we turn to
the partonic initial state. In the parton and hadron cascade model,
PACIAE \cite{PACIAE-intro} the initial partonic state is generated as
follows:

\begin{enumerate}

\item The overlap zone and the number of participant nucleons from the
projectile and target are first calculated geometrically \cite{sa2002}
for an A+A (or A+B) collision, at a given impact parameter.

\item The participant nucleons are distributed randomly inside the overlap
zone, starting from nucleons inside the corresponding nuclear
sphere having an isotropic Woods-Saxon distribution. Nucleons are given beam
momentum, and a particle-list of initial nucleons is constructed.

\item An A+A (A+B) collision is decomposed into nucleon-nucleon (NN)
collision pairs and every one with a collision time calculated by
assuming that the nucleons propagate along straight line trajectories
and interact with the NN inelastic (total) cross sections. Then the
initial NN collision-list is constructed by these NN collision pairs.

The PACIAE model assumes that if a NN collision happens both colliding
nucleons become participants, and eventual occupations of
final particle states are disregarded. These approximations would decrease
the longitudinal fluctuations and angular asymmetries \cite{Yang}.

\item A NN collision pair with the earliest collision time is selected
from the collision list, and the final state of the collision is obtained
by the PYTHIA model with string fragmentation switched-off. Afterwards
the diquarks (anti-diquarks) are broken randomly into quark pairs
(anti-quark pairs), and one obtains a configuration of quarks,
anti-quarks, and gluons, beside a few hadronic remnants for a NN collision.

Although gluons are treated as point like particles, this treatment is
not accurate, as gluons are mediating the interaction among the color
charges and they have significant role in the formation and
hadronization of QGP.  In these transitions, the energy of gluons is connected to the
masses of the hadrons and to the energy of the emitted high energy
photons. We neglect photons in the initial state and so we neglect
pre-equilibrium photon emission also. The detailed treatment
of the gluons, hadron-parton transition and pre-equilibrium emission
would increase the $y_{CM}$ fluctuations. To include these effects would
be overly complicated and not realized in models similar to PACIAE.
Thus, instead we chose to neglect the gluon contribution to $y_{CM}$.
In the present highly approximate treatment, where gluons are treated
as point like classical particles, the inclusion of the gluons would
reduce $y_{CM}$ fluctuations contrary to the physical expectations.

\item Each of the particles (nucleons) travels along straight line
trajectories between two consecutive NN collisions. After the collision
the particle list and collision time list are updated,
the last step and this process are repeated until the NN collision list
becomes empty (the NN collision pairs are exhausted).

\end{enumerate}

The hadron and parton cascade model, PACIAE, includes the
most important geometrical effect of the fluctuation of center of mass
momentum in heavy ion collisions, as the positions of the initial nucleons
are random following the original Woods-Saxon profiles of the projectile
and target nuclei. Then in the overlap region nucleons may collide with
each other according to NN cross section, and those which do not, will
become spectators. This construction provides the participant nucleons,
their positions and momenta, as well as the number of spectators from
the projectile and the target separately. All other effects, which would
influence the $y_{CM}$ fluctuations are neglected. In this way the
model gives a lower limit for the fluctuations of the initial state
CM rapidity.

From the point of view of global collective flow phenomena, we would
have to consider an initial system of particles in
local thermal equilibrium. This system does not contain non-thermalized,
pre-equilibrium emitted particles, jets, high energy direct gammas,
etc. In the present estimate we neglect all these effects, as the
quantitative theoretical estimate of all these effects is exceedingly
difficult, and even the definition of which particles could be considered
belonging to the collective initial state is not settled. These channels
take away considerable energy and momentum from the collective initial
state, so the center of mass rapidity of the collective initial state
will be bigger than the "lower limit" estimate provided by the model
PACIAE.

\subsection{Particle number asymmetries in PACIAE model}

We first estimate the probability distribution of the participant nucleons
suffered at least one nucleon-nucleon collision. Let us have $N\!a$
participant nucleons from the projectile and $N\!b$ from the target. When
$N\!a=N\!b$ the participant matter is symmetric, so the CM momentum and the
CM-rapidity vanish.

\begin{figure}
\vskip -1mm
 \centering
  \includegraphics[width=3.4in]{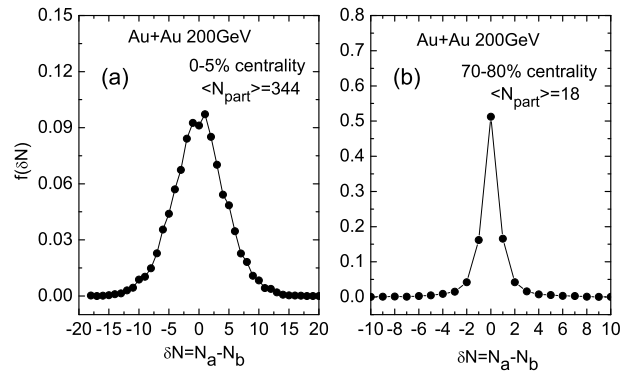}
\vskip -1mm
 \caption{
Initial state fluctuation of the number of extra
nucleons, $\delta N$, in 100+100 A$\cdot$GeV 0-5\% central and
70-80\% peripheral Au+Au collisions in PACIAE model.}
 \label{F_AuAu-dN}
\end{figure}

At a given impact parameter we have a possibility for symmetric
fluctuations when $N\!a=N\!b$ change by equal number of nucleons. This
will not effect the Center of Mass. If we have an asymmetry,
$\delta\! N = N\!a - N\!b$, this leads to a change of the CM-rapidity.

\begin{figure}
\vskip -1mm
 \centering
  \includegraphics[width=3.4in]{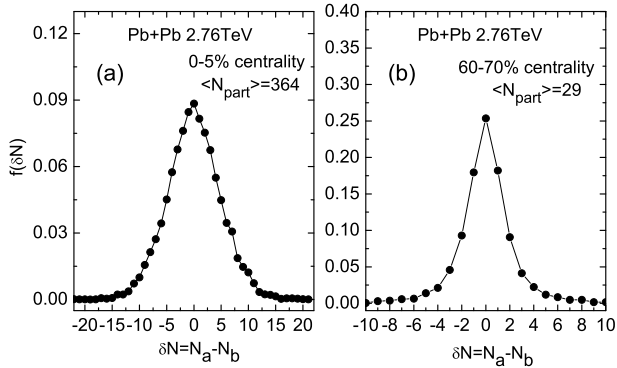}
\vskip -1mm
 \caption{
Initial state fluctuation of the number of extra
nucleons, $\delta\! N$, in 1.38+1.38 A$\cdot$TeV 0-5\% central
and 60-70\% peripheral Pb+Pb collisions in PACIAE model.}
 \label{F_PbPb-dN}
\end{figure}

Taking into account the effect of overlap geometry and of the
nucleon-nucleon cross section, the PACIAE model \cite{PACIAE-intro},
estimates the $\delta\! N$ distribution from $N_{part}$ fluctuations as
presented in Figures \ref{F_AuAu-dN} and \ref{F_PbPb-dN}.

In our model calculations the centrality bins are defined in terms of the
geometrical cross section, $b_{max}^2 \pi = (2 R_A)^2 \pi$,
and for example a centrality bin of 60-70\% corresponds to an impact
parameter range $[ b_i , b_j ]$, such that
$(b_i^2 \pi)/(b_{max}^2 \pi) = 0.6$ and
$(b_j^2 \pi)/(b_{max}^2 \pi) = 0.7$.

As shown in Figure 1 for the central RHIC collisions
$ |\delta N |/ <N_{part}> \approx 1.5 \% $, while for peripheral collisions
it is 5\,\%. In peripheral collisions the longitudinally moving
uppermost and lowermost layers have relatively more particles
than in central collisions, and so the random fluctuations have include
relatively more particles, although the absolute number of particle
asymmetry is less.

In Figure 2 the same results for LHC collisions are 1.4\,\% for central
and 6\,\% for peripheral collisions.
Thus, the relative number fluctuation for central collisions
decreased slightly due to the difference in the number of participants,
while for peripheral collisions the small increase is primarily caused
by the difference in the centrality bin. The small difference indicates
that the relative number fluctuation in peripheral collisions is less
sensitive to centrality bin selection than the absolute numbers.

At higher energy the cross sections are bigger, so both the number of
realized primary-primary collisions and primary-secondary collisions
are bigger. This results in an increase in the participant number in
the same overlap domain.  This leads to the observed fact
that while the absolute numbers are increasing the relative number
fluctuations show a smaller increase.

\subsection{Rapidity fluctuations in PACIAE model}

Let us make a simple estimate: what is the resulting
CM-rapidity fluctuation. The extra nucleons, $\delta\! N$, carry a
longitudinal momentum of
$ \delta p_{z} = \delta N \, m_N \sinh (y_0)$.
The total momentum of the symmetric part,
$ (N\!a+N\!b -|\delta\! N| )$,
of the participant matter
vanishes. We assume a fix impact parameter, $b$ and neglect mass number
fluctuations of the symmetric part of participant matter.
Then we can assume the mass number of the symmetric part to be
$<N_{part}>-<|\delta\! N|>$.
If we assume further that all of the reaction energy is absorbed in the
participant matter and
$<N_{part}>\gg \delta\! N$
then we get
$$
\Delta y_{CM}(\delta N) \approx {\rm artanh}
\left[\frac{\delta N}{<N_{part}>} \tanh(y_0) \right] \, .
$$
Thus, the  CM-rapidity distribution becomes a series of delta functions
according to the $\delta N$-distribution. If we allow for the fluctuation
of the symmetric mass number for a range of impact parameters or a range of
multiplicities, or we allow other channels mentioned above leaking energy
from the initial state the peaks of the CM-rapidity distribution will
be smoothed out.

\begin{figure}[h]
 \centering
  \includegraphics[width=3.4in]{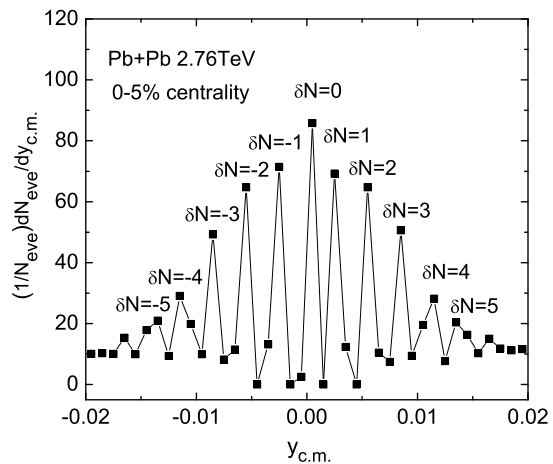}
\vskip -1mm
 \caption{
Initial state CM-rapidity fluctuation in
1.38+1.38 A$\cdot$TeV 0-5\% central Pb+Pb collisions in PACIAE model.
The figure shows that the rapidity change caused by $\delta N$=1, 2,
3, ..., extra nucleons leads to very sharp peak in the CM-rapidity distribution,
each peak corresponding to a given $\delta N$ value. This is
because there is no tightly bound system to absorb energy and momentum
in the model. If not so, the bound system will allow for rapidity
fluctuations at given $\delta N$, making each sharp peak much wider,
and increases the width of the overall $y_{CM}$ distribution.}
\label{F_PbPb-dy}
\end{figure}

Figure \ref{F_PbPb-dy} shows this delta function structure in the
resulting partonic initial state
generated by PACIAE model for 1.38+1.38 A$\cdot$TeV 0-5\% central Pb+Pb
collisions.

The sharp peak structure indicates that all other channels
(pre-equilibrium emissions, etc.) are neglected in our estimate, so
the source of rapidity fluctuations is the momentum of those extra nucleons,
which are not matched in originating from the projectile and the target.

\subsection{CM-fluctuations of different matter components}

In the partonic initial state generated by the PACIAE model a large part of
reaction energy is invested into gluons.
The gluons are treated as classical point like particles just as the
quarks and anti-quarks.  If these gluons were regarded as
a distinct gluon field, then this gluon field might keep the partonic initial
state system more bound and uniform. Then the remaining part
(quarks and anti-quarks) of the partonic initial state
fluctuates stronger.

There are other possibilities, which may increase the CM-rapidity
fluctuation, e.g. pre-equilibrium emission of high energy particles reducing
the energy or mass of the initial state system; considerable kinetic
energy in rotation of the initial state system; etc..

Gluons have an important role in developing collective flow still
in the QGP phase (indicated by the constituent quark number scaling
observed at RHIC). This collective flow at high energies may lead
to a collective rotation \cite{Fluc,CMSS} where a significant
part of the collision energy remains in longitudinal flow, and so
it does not contribute to the transverse mass of the system. This
would lead to a form of collective energy from the gluons, which
leads to increased $y_{CM}$ fluctuations because this energy reduces
the transverse mass of the system. Such, collective effects are not
included in PACIAE, as gluons are treated as classical point like
particles.

The initial state fluctuations of the energetic partonic matter
may be important because the developments of these
components may not be identical, especially at the final FO and hadronization
stages of the reaction. The gluon fields may contribute to forming the
final rest masses of the hadrons, and they may contribute different
amount of thermal and collective kinetic energy to different hadrons
\cite{Zsch2011}.
All effects mentioned above would increase the center of mass rapidity
fluctuation of the initial state, but these are not included in
the PACIAE model we used.

\begin{figure}[h]
 \centering
  \includegraphics[width=3.4in]{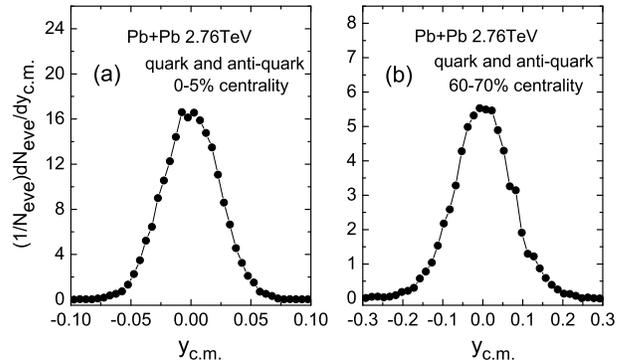}
\vskip -1mm
 \caption{
The CM-rapidity fluctuation of  quarks and anti-quarks in the
initial state
calculated for 1.38+1.38 A$\cdot$TeV, 0-5\% central and 60-70\%
peripheral Pb+Pb collisions by PACIAE model.}
\label{F_PbPb-qdy}
\end{figure}

\begin{figure}[h]
 \centering
\vskip -1mm
  \includegraphics[width=3.4in]{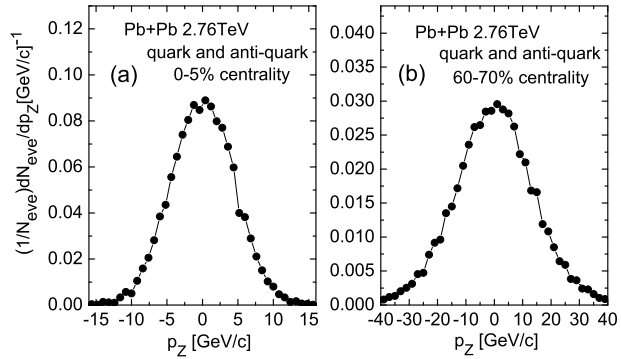}
\vskip -1mm
 \caption{The fluctuation of the CM-longitudinal momentum per participant
nucleon of the quarks and anti-quarks in partonic initial state, i.e. $p_z$
fluctuation calculated for 1.38+1.38 A$\cdot$TeV 0-5\% central and 60-70\%
peripheral Pb+Pb collisions by the PACIAE model.}

 \label{F_PbPb-qpzn}
\end{figure}

Figure \ref{F_PbPb-qdy} gives CM-rapidity fluctuation of the quarks and
anti-quarks in the partonic initial state calculated for 1.38+1.38 A
$\cdot$TeV, 0-5\% central and 60-70\% peripheral Pb+Pb collisions by PACIAE
model. The fact that, the massive gluon field may carry energy and momentum,
makes it possible to incorporate part of the fluctuations.
This enables the model to achieve around a few times larger CM-rapidity
fluctuations than without a flexibly moving massive gluon field as one can
see in comparing Fig. \ref{F_PbPb-dy} with Fig. \ref{F_PbPb-qdy}.
Figure \ref{F_PbPb-qpzn} gives the fluctuation of the CM-longitudinal
momentum per participant nucleon of the quarks and anti-quarks in the
partonic initial state, i.e. $p_z$ fluctuation.

In the PACIAE model calculations above, nearly 57.6\% of the
total collision energy is shared by the quarks and anti-quarks and 42.4\% by
the gluons in the 60-70\% centrality Pb+Pb collisions. These values are
57.9\% and 42.1\% for quarks and anti-quarks and gluons, respectively, in
the 0-5\% central Pb+Pb collisions. So, how gluons are treated is an
important issue.

%
%
%
%
%

\section{Conclusions}\label{Conc}

Initial state fluctuations were analyzed in the PACIAE model, with
particular attention to the CM-rapidity fluctuations.
It was found that in central collisions the longitudinal asymmetry,
arising from different number of projectile and target participants,
in longitudinal momentum is around 1.5\% only, while for peripheral reactions
it can reach $\pm 5 - 6\%$ (see Figs. \ref{F_AuAu-dN} and \ref{F_PbPb-dN}).
In central collisions the CM-rapidity fluctuations arise from this
longitudinal asymmetry is not large in the PACIAE
model as indicated by Figure \ref{F_PbPb-dy}.

We can see in Fig. \ref{F_PbPb-qdy} that the arising CM-rapidity fluctuation
is around $\pm 0.03$ units for central collisions and around $\pm 0.1$ units
in peripheral ones. These are about $5-10$ times
smaller than the assumptions used in ref. \cite{CMSS}, and this would
result in less reduction of the original $v_1(y)$ calculations. On the
other hand, the PACIAE estimates can be considered as conservative
lower limits of the $y_{CM}$ fluctuations, so the measured $y_{CM}$
fluctuations may exceed these values. In the present formulation
of PACIAE, with point like gluons, the gluon contribution would
decrease the CM-rapidity fluctuations (cf. Fig. \ref{F_PbPb-dy}).

In the PACIAE partonic initial state study above, we
do not include the pre-equilibrium emission, the collective
effects as e.g. rotation, and the formation of excited
intermediate states. These could lead to the increase of
CM-rapidity fluctuations. The developing collective flow
may increase and decrease fluctuations, depending on the
quantitative details of the developing flow pattern. The structure
of the collective flow will be detected at the end of the
reaction, but this pattern develops from the initial
state in the QGP phase where the gluon component is essential.
The collective flow has both transverse and longitudinal components.
The pre-collision initial state has exclusively longitudinal
collective motion. At the time point of strongest stopping, this
longitudinal flow energy is reduced to about 30\% of the initial
value, while in average at the end of the reaction the longitudinal
and transverse energy has about $50-50\%$ share\cite{LBLP}. Soft EoS (like QGP) and
collective rotation may increase the share of longitudinal flow
energy. The increased longitudinal energy (especially from rotation)
and the projectile/target participant asymmetry may in itself
contribute to direct increased longitudinal fluctuation.

The share of longitudinal and transverse flow energies also
influence the transverse mass of the system, which indirectly
contributes to longitudinal fluctuations. The transverse
part of the flow energy increases the transverse mass, while the
longitudinal part reduces it. Larger transverse mass reduces
the $y_{CM}$ fluctuations. We know that with increasing beam energy
the collective flow became more energetic and it is the most
dominant phenomenon at LHC energies. This arises from the
initial energy and momentum distribution, including the
gluon components, as these are necessary for the development
of the large collective flow processes.

The PACIAE model with point like gluons has less ability to
incorporate these collective flow effects, and about two-thirds of the
available energy will contribute to the transverse mass, while
no direct longitudinal flow fluctuation will develop from
the initial state asymmetries. Thus, PACIAE with point like
gluons underestimates the $y_{CM}$ fluctuations.

The initial state longitudinal fluctuations are essential for the
analysis of the directed flow, as these fluctuations have significant
effect on the measurable $v_1$-flow \cite{CMSS}.
The present situation regarding the directed flow is rather complex
as at RHIC and LHC energies that, the observed collective $v_1$ flow is
rather weak, $| v_1 | \le 0.001$ at $\eta = 0.8$,
so the $v_1$-flow from  the initial state fluctuations may exceed the
global collective $v_1$ flow. Thus, the evaluation of $v_1(p_t)$
at low momenta and low rapidities is a complex problem, where the
two processes are interacting \cite{QM-v1}.
The event-by-event longitudinal fluctuations may be important in
the assessment and separation of the global directed flow and the
directed flow arising from the initial state random fluctuations.
\vskip -1mm

\section*{Acknowledgements}\label{Ack}

This work was supported by the National Natural Science Foundation of China
under Grants No. 11075217, 11047142, 10975062 and the 111 project
of the foreign expert bureau of China.
L.P. Csernai thanks for the kind hospitality of the Institute of Particle
Physics of the Huazhong Normal University, where part of this work
was done.


\begin{thebibliography}{99}

\bibitem{ALICE-Flow1}
  K. Aamodt et al., (ALICE
  Collaboration) Phys. Rev. Lett. 105, 252302 (2010).

\bibitem{Son}
  P.K. Kovtun, D.T. Son and A.O.
  Starinets, Phys. Rev. Lett. 94, 111601 (2005).

\bibitem{CKM}
  L.P.Csernai, J.I.Kapusta, L.D.{McLerran}.
  Phys. Rev. Lett. 97, 152303--4 (2006).

\bibitem{Hybrid}
  S.A. Bass, A. Dumitru, M. Bleicher, L. Bravina, E. Zabrodin,
  H. Stocker, and W. Greiner, Phys. Rev. C 60,  021902 (1999);
  T. Hirano, P. Huovinen, Y. Nara, arXiv:1012.3955v1 [nucl-th];
  H. Petersen, G.Y. Qin, S.A. Bass, and B. Muller, Phys.
  Rev. C 82, 041901 (2010);
  B. Bauchle and M. Bleicher, Phys. Rev. C81, 044904 (2010).

\bibitem{Fluc}
  H. Petersen, Ch. Coleman-Smith, S.A. Bass, R. Wolpert,
  J. Phys. G 38, 045102 (2011);
  F.G. Gardim, F. Grassi, Y. Hama, M. Luzum, J.Y. Ollitrault,
  Phys. Rev. C 83, 064901 (2011);
  P.~Bozek, W.~Broniowski, J.~Moreira,
Phys.\ Rev.\  C83, 034911 (2011);
P.~Bozek, I. Wyskiel,
Phys.\ Rev.\  C81, 054902 (2010).



\bibitem{Qin10}
  G.Y. Qin, H. Petersen, S.A. Bass and
  B. Muller, Phys. Rev. C 82, 064903 (2010);

\bibitem{Pet11}
  H. Petersen, V. Bhattacharya,
  S.A. Bass and C. Greiner, arXive: nucl-th/1105.0340v1]

\bibitem{Grassi}
   F.G. Gardim, F. Grassi, Y. Hama, M. Luzum, J.Y. Ollitrault,
   arXiv:1103.4605 [nucl-th]

\bibitem{CK92}
  L.P.~Csernai, and J.I.~Kapusta, Phys.~Rev.~Lett.~{\bf 69}, 737 (1992); and
  L.P.~Csernai, and J.I.~Kapusta, Phys.~Rev.~D {\bf 46}, 1379 (1992).

\bibitem{CMSS}
  L.P. Csernai, V.K. Magas, H. St{\"o}cker, D.D. Strottman,
  arXiv: 1101.3451v3 [nucl-th].


\bibitem{M2001-2}
  V.K.~Magas, L.P.~Csernai, and D.D.
  Strottman, Phys. Rev. C 64 (2001) 014901;  and
  V.K.~Magas, L.P.~Csernai, and D.
  Strottman, Nucl. Phys. A 712 (2002) 167.

\bibitem{hydro1}
  L.P. Csernai, Y. Cheng, V.K. Magas, I.N. Mishustin, and D. Strottman,
  Nucl. Phys. A 834, 261c (2010).


\bibitem{PACIAE-intro}
  Ben-Hao Sa, Xiao-Mei Li, Shou-Yang Hu, Shou-Ping Li, Jing Feng, and
  Dai-Mei Zhou,
  Phys. Rev. C 75 054912 (2007); and
  Ben-Hao Sa, Dai-Mei Zhou, Yu-Liang Yan, Xiao-Mei Li, Seng-Qin Feng,
  Bao-Guo Dong, and Xu Cai,
  arXiv: 1104.1238v1/nucl-th

\bibitem{sa2002}
Ben-Hao Sa, A. Bonasera, An Tai, and Dai-Mei Zhou, Phys. Lett. B 537, 268
(2002).

\bibitem{Yang}
C.B. Chiu, R.C. Hwa, C.B. Yang, Phys. Rev. C 78, 044903 (2008).

\bibitem{Zsch2011}
S. Zschocke, S. Horvat, I.N. Mishustin, and L.P. Csernai,
  Phys. Rev. C 83, 044903 (2011).

\bibitem{LBLP}
L.Bravina, L.P Csernai, P.Levai, D. Strottman,
Phys. Rev. C 50, 2161 (1994)

\bibitem{QM-v1}
  Contributions of R. Snellings, I. Selyuzhenkov and G. Eyyubova et. al.
  (all for the ALICE collaboration) at the Quark Matter 2011 Conf.
  Annecy France, May 22-28, 2011, to be published in the proceedings.


\end{thebibliography}
\end{document}